\documentclass[runningheads,citeauthoryear]{apinv}
\usepackage{epsfig,cite,color,graphics}
\usepackage{marvosym} %For astronomical symbols % % %
\usepackage[utf8]{inputenc}

\begin{document}

\title{Thermodynamics of fluid elements in the context of saturated isothermal turbulence in molecular clouds}
\titlerunning{Thermodynamics of fluid elements}
\author{Sava Donkov\inst{1}, Ivan Zh. Stefanov\inst{2}, Todor V. Veltchev\inst{3}}
\authorrunning{S. Donkov, I. Stefanov, T. Veltchev}
\tocauthor{Sava Donkov, Ivan Zh. Stefanov, Todor V. Veltchev} 
% Command tocautor{} is used by the Latex to give author names 
% to the Contents of the volume (automatically generated)
\institute{Institute of Astronomy and NAO, Bulgarian Academy of Sciences, 72 Tzarigradsko Chausee Blvd., 1784 Sofia, Bulgaria
	\and Department of Applied Physics, Faculty of Applied Mathematics, Technical University-Sofia, 8 Kliment Ohridski Blvd., Sofia 1000, Bulgaria
	\and Faculty of Physics, University of Sofia, 5 James Bourchier Blvd., 1164 Sofia, Bulgaria   \newline
	\email{sddonkov@astro.bas.bg}    }
\papertype{Submitted on xx.xx.xxxx; Accepted on xx.xx.xxxx}	
% Papertype can be "Research report", "Review", "Invited lecture", "Conference talk", 
% "Conference poster", "Lecture at scientific seminar", "Summary of dissertation",  etc.
\maketitle

\begin{abstract}

The presented paper is an attempt to investigate the dynamical states of an hydrodynamical isothermal turbulent self-gravitating system using some powerful tools of the classical thermodynamics. Our main assumption, inspired by the work of Keto et al. (2020), is that turbulent kinetic energy can be substituted for the macro-temperature of chaotic motion of fluid elements. As a proper sample for our system we use a model of turbulent self-gravitating isothermal molecular cloud which is at final stages of its life-cycle, when the dynamics is nearly in steady state. Starting from this point, we write down the internal energy for a physically small cloud's volume, and then using the first principle of thermodynamics obtain in explicit form the entropy, free energy, and Gibbs potential for this volume. Setting fiducial boundary conditions for the latter system (small volume) we explore its stability as a grand canonical ensemble. Searching for extrema of the Gibbs potential we obtain conditions for its minimum, which corresponds to a stable dynamical state of hydrodynamical system. This result demonstrates the ability of our novel approach.

\end{abstract}
\keywords{thermodynamics hydrodynamics stability molecular clouds}

\section*{Introduction}
\label{Introduction}

In this work we set ourself a task to adopt the notions and methods of classical thermodynamics as tools of describing the dynamical states of one hydrodynamical isothermal turbulent self-gravitating system. In particular, we are interested in equilibrium states. As a model for this purpose we use sample of one typical interstellar molecular cloud (MC), at the last stages of its life cycle, when the cloud is at nearly steady state and hence the hydrodynamical system is closed to dynamical equilibrium.

Molecular clouds are the places of star formation in our Galaxy which highlights their importance for galactic evolution (Hennebelle \& Falgarone, 2012; Klessen \& Glover, 2016). They are gaseous structures of irregular shape and consist mostly of cold ($T\sim 10 - 30$~K, Ferriere 2001) molecular  gas and a small amount of interstellar dust ($\sim 1\%$). Their physics is complex and governed by the interplay between gravity, accretion from the surrounding medium, supersonic turbulence and magnetic fields, with nearly isothermal equation of state (Hennebelle \& Falgarone, 2012; Klessen \& Glover, 2016). The feedback of new-born stars and supernovae explosions which eventually disrupt the parental cloud complete the picture.

MCs own fractal structure in a large interval of spatial scales – 0.001 pc $\leq l \leq$ 100 pc , which is thought to be, also, the inertial range for turbulence there (Elmegreen, 1997; Elmegreen \& Scalo, 2004; Hennebelle \& Falgarone, 2012; Klessen \& Glover, 2016). Moreover the density contrast between large and small substructures covers several orders of magnitude. At scales $l \sim$ 100 pc the density $n$ is about $10^2$ cm$^{-3}$, while at scales of pre-stellar cores ($l \leq$ 0.1 pc) $n$ is of order $10^5$ cm$^{-3}$ and more. If we look at more dense substructures, we arrive at scales of pre- and protostellar cores.

We model our cloud as a cold isothermal gaseous structure embedded in an extended, but not infinite, medium consisting of atomic hydrogen which serves as a material and thermal reservoir for the MC. A phase transition takes place at the boundary between the cloud and its surroundings, i.e. when density and temperature of the inflowing gas change. We assume that there exists an inertial range of scales for turbulence in the cloud which spans from its boundary down to about the sizes of protostellar cores. This turbulence is presumed to be fully developed and saturated, and also locally isotropic and homogeneous. We regard the gravitational field in a fixed small volume in the cloud as caused by self-gravity of the cloud and gravity of the surrounding medium. Our main assumption is that the turbulent kinetic energy can be locally substituted for the macro-temperature of chaotic motion of the fluid elements. Indeed, we treat this motion as purely chaotic (locally) and hence the fluid elements act as particles of a perfect gas. The idea was originally suggested by Keto et al. (2020) with the purpose to investigate instability and fragmentation of star-forming clouds by use of turbulent-entropic instability as a thermodynamic tool. In the present work, starting from the above assumption, we write down the internal energy, for a physically small volume, of a turbulent self-gravitating and isothermal MC. The model is presented in more details in Section 1. In Section 2, using the first principle of thermodynamics, we obtain the explicit expressions for entropy, free energy and Gibbs potential of the macro-gas in regarded small volume. Setting proper boundary conditions for the latter system, we study in Section 3 its stability as a grand canonical ensemble and obtain conditions for a minimum of the Gibbs potential, which corresponds to a stable dynamical state of hydrodynamical system. This result demonstrates the ability of our novel approach. In Section 4 we list several possible caveats against the model and comment on them briefly. This Section ends with a short conclusion.

\section*{1. Setup of the model}
\label{Sec-set up of the model}

We consider a cloud of molecular gas, which is turbulent and isothermal, i.e. $T= {\rm const}$ (Ferriere 2001). The turbulence is fully developed and saturated. There exist an inertial range of scales: $l_{\rm d} \leq l \leq l_{\rm up}$ (Elmegreen \& Scalo, 2004; Hennebelle \& Falgarone, 2012; Klessen \& Glover, 2016), where $l_{\rm d}$ is the turbulence dissipation scale and $l_{\rm up}$ is the upper limit of the range. We suppose that the outer size of the cloud $l_{\rm c}$ obeys the conditions: $l_{\rm c} \leq l_{\rm up}$ and $l_{\rm d} \ll l_{\rm c}$.

The turbulence, locally, is homogeneous and isotropic (Elmegreen \& Scalo, 2004; Hennebelle \& Falgarone, 2012; Klessen \& Glover, 2016) and the motion of fluid elements is purely chaotic. Therefore this local motion can be modelled as a perfect gas of fluid elements with a macro-temperature $\theta$, expressed through the following equation:

\begin{equation}
	\label{equ-intro theta}
	\frac{1}{2} m \sigma(l)^2 \equiv \frac{3}{2} \kappa\theta(l)~,
\end{equation}
where $\sigma(l) = u_0 l^{\beta}$ is the 3D velocity dispersion of turbulent motion of fluid elements at scale $l$, $u_o\sim1$ is a normalizing coefficient, $0<\beta\leq1$ is a scaling exponent (Larson, 1981; Padoan et al., 2006; Kritsuk et al., 2007), $m$ is the mass of fluid elements and $\kappa$ is the Boltzmann constant.

The molecular cloud (MC) is embedded in a very large, but not infinite, medium which consists of atomic hydrogen and causes a constant gravitational potential $\varphi_{\rm m} = {\rm const}$ in the cloud volume. The modelled MC can exchange particles/fluid elements with the surrounding medium. Also, when the gas flows through the cloud boundary, it becomes molecular, its Kelvin temperature becomes equal to the cloud's temperature (gas is cooling) and its density becomes equal to the cloud's boundary density $\rho_{\rm c}$. Only its pressure does not change. In other words, a phase transition occurs at the cloud boundary \footnote{In the theory of interstellar medium the transition from atomic to molecular gas is regarded as a first order phase transition with only the gas pressure stays unchanged. During the process gas releases constant energy per volume, because the temperature of molecular hydrogen is about two orders of magnitude lower than this of atomic hydrogen.}.

We assume that the cloud is at the last stages of its evolution and hence, dynamically, it is in a nearly steady state (Slyz et al., 2005; Kritsuk, Norman \& Wagner, 2011; Girichidis et al. 2014; Schneider et al., 2015a; Schneider et al., 2015b; Schneider et al., 2015c; Schneider et al., 2016; Veltchev et al., 2019).

At any cloud scale $l \leq l_{\rm c}$ we consider a physically small volume $V_0 \ll l^3$, which is, although, sufficiently large in regard to the chaotic turbulent motion of fluid elements. This volume is presumably constant at any cloud scale in the inertial range. Therefore, if the condition $V_0 \ll l_{\rm d}^3$ is satisfied, it is true for all cloud scales. The gravitational potential in this volume reads: $\varphi(l) = \varphi_{\rm s}(l) + \varphi_{\rm m}$, where $\varphi_{\rm s}(l)$ is caused by the self-gravity of the cloud. We denote with $n(l)$ the number density of macro-gas at scale $l$ and suppose that $n(l) = {\rm const}$ in the volume $V_0$, i.e. this volume is homogeneous.

We assume that some quantities depend solely on scale $l$. This is strictly true only in case of spherical symmetry and isotropy, but it is adopted here for simplicity of the considerations, which is justified as a first step in the development of our model.

Now we write down the formula for internal energy of the macro-gas in the volume $V_0$:

\begin{equation}
	\label{equ-int en}
	\varepsilon(l) = \frac{3}{2} N(l) \kappa \theta(l) + \frac{3}{2} \frac{m}{m_0} N(l) \kappa T + m N(l) \varphi(l) ~,
\end{equation}
where $m_0$ is the averaged molecular mass of the gas and $N(l)$ is the number of fluid elements in volume $V_0$. Hereafter, the explicit dependence of $l$ is omitted in the equations for simplicity, unless it is important to be pointed out.

\section*{2. Equations}
\label{Sec-equations}

In this Section we obtain the explicit form of several basic thermodynamic potentials (namely: entropy, free energy, and Gibbs potential) which are important for the further considerations.

From equation (\ref{equ-int en}) one can see that the internal energy is a function of two thermodynamic variables: $\varepsilon = \varepsilon(\theta,N)$, as it is natural for an homogeneous thermodynamic system. Then the total differential ${\rm d}\varepsilon(\theta,N)$ reads as follows:

\begin{equation}
	\label{equ-diff int en}
	{\rm d}\varepsilon = \frac{3}{2} N \kappa {\rm d}\theta + \bigg( \frac{3}{2}\kappa\theta + \frac{3}{2} \frac{m}{m_0}\kappa T + m\varphi \bigg){\rm d}N ~.
\end{equation}

On the other hand, from the first principle of thermodynamics, one gets:

\begin{equation}
	\label{equ-first principle}
	{\rm d}\varepsilon = \theta {\rm d}s - P{\rm d}V_0 + \mu {\rm d}N ~,
\end{equation}
where $s$ is the entropy in volume $V_0$, $P$ is the pressure of macro-gas, and $\mu$ is the chemical potential.

In the case $N = {\rm const}$ and taking into account the constant volume $V_0$: $P{\rm d}V_0=0$, we derive from equations (\ref{equ-diff int en}) and (\ref{equ-first principle}) the following expression:

$${\rm d}s = \frac{3}{2}\frac{N\kappa}{\theta}{\rm d}\theta ~.$$

Integrating it in the limits from the dissipation scale $l_{\rm d}$ to the scale in consideration $l$, one obtains:

\begin{equation}
	\label{equ-entropy}
	s(\theta,N) = \frac{3}{2} N\kappa \ln(\theta/\theta_{\rm d}) ~,
\end{equation}
where we presume that $s_{\rm d} = s(\theta_{\rm d},N) = 0$, which plays a role of the third principle of thermodynamics of fluid elements and is a natural assumption, because turbulence dissipates at scale $l_{\rm d}$. 

By use of equations (\ref{equ-int en}) and (\ref{equ-entropy}) one can derive the explicit form of free energy: $f(\theta,N) = \varepsilon(\theta,N) - \theta s(\theta,N)$, and the respective formula is as follows:

\begin{equation}
	\label{equ-free energy}
	f(\theta,N) = \frac{3}{2} N\kappa\theta [1 - \ln(\theta/\theta_{\rm d})] + \frac{3}{2} \frac{m}{m_0} N\kappa T + mN\varphi ~.
\end{equation}

Now the chemical potential $\mu$ reads:

\begin{equation}
	\label{equ-chem potential}
	\mu = \bigg( \frac{\partial f}{\partial N} \bigg)_{\theta} = \frac{3}{2} \kappa\theta [1 - \ln(\theta/\theta_{\rm d})] + \frac{3}{2} \frac{m}{m_0} \kappa T + m\varphi ~.
\end{equation}

Finally we obtain the explicit formula for Gibbs potential (hereafter, Gibbs potential or Gibbs energy): $g(\theta,N) = \varepsilon(\theta,N) - \theta s(\theta,N) + PV_0$. The latter reads:

\begin{equation}
	\label{equ-gibbs potential}
	g(\theta,N) = \frac{3}{2} N\kappa\theta [(5/3) - \ln(\theta/\theta_{\rm d})] + \frac{3}{2} \frac{m}{m_0} N\kappa T + mN\varphi ~,
\end{equation}
where $PV_0$ is replaced according to the equation of state of the perfect gas of fluid elements: $PV_0=N\kappa\theta$.

\section*{3. Stability analysis of the system}
\label{Sec-stability analysis}

In this section we aim to explore whether the MC regarded as a macro-gas of fluid elements can be in a stable dynamical state. To perform this analysis, one needs to determine the boundary conditions at which the cloud resides. According to the recent knowledge about interstellar medium, a good approximation might be to consider the modelled cloud at fixed macro-temperature $\theta_0$ and pressure $P_0$ of the surrounding medium (see Hennebelle \& Falgarone, 2012; Klessen \& Glover, 2016). We also presume that the number $N$ of the fluid elements in the cloud is constant. It is worth to note here, that the fixed conditions for macro-gas determine not the microstates but rather the macrostates of the gas. Thus, studying the states of the macro-gas, we explore indeed its hydrodynamical states.

We apply the same conditions to an arbitrary small volume $V_0$, at a scale $l$ in the cloud interior. The latter means that the considered small volume $V_0$ is regarded as a grand canonical ensemble -- at fixed macro-temperature, pressure and number of fluid elements. The relevant thermodynamic potential is the Gibbs energy $g(\theta,N,V_0)$. In the following treatment we explore whether this quantity has extrema and what of kind they are (see Reif 1965).

Let us consider the Gibbs energy in a nonequilibrium form:

\begin{equation}
	\label{equ-gibbs potential nonequ}
	g(\theta,N,V_0)=\varepsilon(\theta,N) - \theta_0 s(\theta,N) + P_0 V_0 ~.
\end{equation}

We take partial derivative of $g$ in regard to $\theta$ and equate it to zero. Making use of equations (\ref{equ-int en}) and (\ref{equ-entropy}), it reads:

\begin{equation}
	\label{equ-dg/dtheta}
	\bigg( \frac{\partial g}{\partial \theta} \bigg)_{N,V_0} = \bigg( \frac{\partial \varepsilon}{\partial \theta} \bigg)_{N,V_0} - \theta_0 \bigg( \frac{\partial s}{\partial \theta} \bigg)_{N,V_0} +0 = \frac{3}{2} N\kappa \bigg[ 1 - \frac{\theta_0}{\theta} \bigg] = 0 ~,
\end{equation}
which leads to the first condition for a possible extremum: $\theta = \theta_0$. In other words, the macro-temperature of the volume $V_0$ must be equal to the macro-temperature of the surrounding medium.

To study the condition for pressure, let us take the partial derivative of $g$ in regard to $V_0$. Using the equation of the first principle (\ref{equ-first principle}) and the one for the entropy: (\ref{equ-entropy}), one obtains easily:

\begin{equation}
	\label{equ-dg/dV_0}
	\bigg( \frac{\partial g}{\partial V_0} \bigg)_{\theta,N} = \bigg( \frac{\partial \varepsilon}{\partial V_0} \bigg)_{\theta,N} - \theta_0 \bigg( \frac{\partial s}{\partial V_0} \bigg)_{\theta,N} + P_0 = -P - 0 + P_0 = 0 ~,
\end{equation}
from which follows that the pressure $P$ in the volume $V_0$ must be equal to the pressure $P_0$ of its surroundings.

In order to conclude whether an extremum exists and of what kind, one needs of second partial derivatives of $g$, which are the elements of the corresponding functional determinant $D$. The latter are not difficult to be obtained; calculated at $\theta=\theta_0$ and $P=P_0$, they read:

\begin{eqnarray}
	\label{equ-second patrial der}
	\bigg( \frac{\partial^2 g}{\partial \theta^2} \bigg)_{N,V_0} = \frac{3}{2} \frac{N\kappa}{\theta_0} ~~;~~ \bigg( \frac{\partial^2 g}{\partial V_0^2} \bigg)_{\theta,N} = - \bigg( \frac{\partial P}{\partial V_0} \bigg)_{\theta,N} = \frac{N\kappa\theta_0}{V_0^2} ~~; \nonumber\\
    \frac{\partial^2 g}{\partial \theta \partial V_0} = - \bigg( \frac{\partial P}{\partial \theta} \bigg)_{N,V_0} = - \frac{N\kappa}{V_0} ~~;~~ \frac{\partial^2 g}{\partial V_0 \partial \theta} = 0 ~~.
\end{eqnarray}

Finally we obtain that $D = (3/2) (N\kappa/V_0)^2 > 0$. The latter means that the Gibbs energy has a minimum at $\theta=\theta_0$ and $P=P_0$. Hence, the regarded small volume $V_0$ is in a stable dynamical state.

What can one conclude about the whole cloud? The obtained above conditions are valid for an arbitrary located small volume $V_0$, at an arbitrary scale $l$ belonging to the inertial range. These conditions are quite natural, because, due to the small size of the volume $V_0$ in regard to the scale in considerations, one expects that macro-temperature $\theta$ and pressure $P$ do not change substantially in the vicinity of $V_0$, whose linear size $V_0^{1/3}$ is much less than $l$. Therefore the medium in the cloud is locally dynamically stable, and if the macro-temperature and pressure change continuously through the fluid (which is the case for the scales within the inertial range), then large parcels of the cloud should be stable. The latter conclusion would be valid for the whole cloud if the conditions for $\theta$ and $P$ change continuously through its outer boundary.

\subsection*{4. Discussion and conclusion }
\label{Sec-discussion and conclusion}

In the presented work we attempt to adopt some notions and tools of thermodynamics to study the dynamical states of an hydrodynamical isothermal turbulent and self-gravitating system realized as a model of molecular cloud. This aim can face several caveats as follows.

The first caveat concerns the basic assumption that the turbulent kinetic energy can be substituted for the macro-temperature of the chaotic motion of fluid elements. This is justifiable only locally due to the existence of scales within the presumed inertial range of turbulence (Elmegreen \& Scalo, 2004; Klessen \& Glover, 2016). Hence, if the turbulent velocity fluctuations scale (Kritsuk et al., 2007) then the macro-temperature will also scale, according to equation (\ref{equ-intro theta}). Then a certain temperature $\theta$ can be regarded only locally and not to be ascribed to the whole cloud. This problem is avoided in our work by the consideration of a physically small volume $V_0$ (whose size is much smaller than the scale $l$, at which it resides) which is indeed the studied thermodynamic system.

The second caveat may arise in view of the homogeneity of the system. It is well known that a simple thermodynamic system must be homogeneous (Reif, 1965). Obviously the whole cloud or large parts of it do not satisfy this condition, due to the discovered scaling law for density (Larson, 1989; Kritsuk et al., 2007). But the regarded small volume $V_0$ can be considered as homogeneous, because $V_0^{1/3}\ll l$.

The third caveat may stem from the way self-gravity is accounted for in the equations for different energies (internal energy, free energy, and Gibbs energy). The latter usually causes an issue since self-gravity depends on the square of mass and hence energy is not additive in regard to mass (about additivity of energies see, for example, Reif, 1965). In our treatment we account for self-gravity as an external field, caused by the whole scale $l$ to which the volume $V_0$ belongs (see Section \ref{Sec-set up of the model}). Therefore the equations for energies remain additive.

The last caveat we mention here concerns the dynamics of the studied hydrodynamical system: the modelled cloud. The approach in this work is based on equilibrium thermodynamics. Of course, the notion of temperature is not applicable if the system is not in equilibrium. MCs are systems that are not in hydrodynamical equilibrium during most of their life-time (V\'azquez-Semadeni et al., 2019). However, at the last stages of their life-cycle they are in a nearly steady state (Slyz et al., 2005; Kritsuk, Norman \& Wagner, 2011; Girichidis et al. 2014; Schneider et al., 2015a; Schneider et al., 2015b; Schneider et al., 2015c; Schneider et al., 2016; Veltchev et al., 2019). Thus equilibrium thermodynamics is appropriate to describe this last period of their life.

In conclusion, we argue that our novel approach, inspired by the work of Keto at al. (2020), shows the ability of the classical thermodynamics to provide a fiducial description of equilibrium dynamical states of a hydrodynamical isothermal turbulent self-gravitating system, represented here by a model of molecular cloud. Despite of the used several approximations in regard to the presented physical picture, we consider this work as a sensible step in this direction.

{\it Acknowledgement:} S.D. acknowledges support by the Deutsche Forschungsgemeinschaft (DFG) under grant KL 1358/20-3. T.V. acknowledges support by the Deutsche Forschungsgemeinschaft (DFG) under grant KL 1358/20-3 and additional funding from the Ministry of Education and Science of the Republic of Bulgaria, National RI Roadmap Project DO1-176/29.07.2022.

\end{document}